\documentclass[doublecol]{epl2} 

\title{Conductance Switching and Inhomogeneous Field Melting 
in the Charge Ordered Manganites}

\shorttitle{Field melting in CO manganites}

\author{Anamitra Mukherjee, Kalpataru Pradhan \and Pinaki Majumdar}
\shortauthor{Anamitra Mukherjee, Kalpataru Pradhan and Pinaki Majumdar}

\institute{                    
\inst{} Harish-Chandra  Research Institute,
Chhatnag Road, Jhusi, Allahabad 211019, India
}


\abstract{
The field induced switching of conductance in the charge ordered half-doped
manganites is controlled by the combination of  metastability, an inhomogeneous
high field state, and cation disorder. We study this non-equilibrium problem
via real space Monte Carlo on a disordered strong coupling model appropriate to
the manganites. We reproduce the variation  of the switching fields with the
mean ionic radius $r_A$ and cation disorder $\sigma_A$, and demonstrate how the
experimental features arise from the proximity of several phases in the Landau
free energy landscape. Our prediction for the field melted state is consistent
with a growing body of experimental evidence.
}

\begin{document}

\maketitle

The manganites owe their fame to the `colossal magnetoresistance' effect,
whereby an applied magnetic field strongly suppresses the resistance of the
material. The most extreme version of this happens when the field can actually
induce an insulator-metal transition (IMT). This occurs in the low bandwidth
half doped manganites, {\it e.g}, Pr$_{0.5}$Ca$_{0.5}$MnO$_3$, that are
typically charge ordered (CO) insulators with CE magnetic
order\cite{mang-book,tok-rev,bicr-prl}.  The CE-CO-I state is transformed to a
ferromagnetic metal (FM-M) by an applied field through a first order phase
transition. The  magnetic field induced conversion of the CE order to FM,
and the crucial resistive switching, involve several subtleties which are only
gradually being
appreciated\cite{respaud2000,tokura96,kuwahara97,tomioka95,wu,pram1,pram2,chad,trokiner,kawano97,kirste2003}.
These include the surprising difference between the melting fields in two
lanthanide (Ln) families, namely the Ca \cite{respaud2000} and the Sr
\cite{tokura96,kuwahara97} families; the smallness of the field melting
energy scales \cite{tomioka95,respaud2000,kuwahara97,tokura96}; and the \textit{
spatial character} of field induced melting and its relation to structural
disorder\cite{wu,pram1,pram2,chad,trokiner}. We expand on these issues
below.

(i)~The phase boundary between the CE and FM phase is first order, so 
metastable
states exist on either side. This leads to hysteresis, and 
the switching fields
are decided by {\it non-equilibrium effects}, not free energy balance.
(ii)~While trapping into metastable states is expected in a first order
transition, an applied field leads to {\it new equilibrium states}
which are not simple continuation of the zero field state. In particular, it is 
increasingly apparent that the field induced conducting state is an {\it
inhomogeneous} FM-M, a percolative metal  with a finite volume fraction of
charge ordered regions\cite{wu,pram1,pram2,chad,trokiner}. Finally, (iii)~the
field induced switching and inhomogeneous metallicity play out on a structurally
disordered background. One expects the CO-I state to be more stable in narrow
band systems. While this is true in the Ca based manganites\cite{respaud2000},
where the cation size mismatch $\sigma_A$ is small, it is completely opposite in
the Sr based systems\cite{tokura96,kuwahara97}. The more insulating system can
be driven metallic {\it easier} in the presence of disorder!

These three actors, metastability, field induced inhomogeneity, and pinning
disorder, enrich and complicate the field melting problem. There is no
understanding of their interplay in the field induced IMT, and the
non-equilibrium physics of the manganites. 

We study a coupled electron-spin-lattice model \cite{kp-am-pm1} appropriate to
the manganites, in two dimensions (2D). For parameters where the FM-M and
CE-CO-I phases are in proximity, we discover the following. (i)~In the `clean'
limit, the forward (CE to FM) and reverse (FM to CE) switching fields, $h_c^+$
and $h_c^-$ respectively, increase rapidly with decreasing bandwidth (BW), and
below a  threshold BW the field cannot melt the CO. This explains the results in
the `clean' Ca based manganites. (ii)~The switching fields are dramatically
affected by disorder: $h_c^{\pm} \rightarrow 0$ as the bandwidth reduces, as in
the Sr manganites, even though the CO is notionally stronger.
(iii)~The field melted state in the narrow band case is 
inhomogeneous even
without disorder - at finite field the system exists as a patchwork 
of metallic
and CO regions, and the `metallicity' is of percolative origin. 
Recent
experiments have indeed verified coexistence of FM-M and CO-I, 
{\it e.g}, in
La$_{0.5}$Ca$_{0.5}$MnO$_3$\cite{chad} and
Pr$_{0.5}$Sr$_{0.5}$MnO$_3$\cite{pram1,pram2}, and the phase volumes 
are tunable depending on the field cooling protocol.

We consider the field response of a model $H_0$ that we had 
studied  earlier
\cite{kp-am-pm1} to capture phase competition.  
Our model is:
\begin{eqnarray}
H &=& H_0 - \mu {\hat N} + \sum_i \epsilon_i n_i - h\sum_i S_{iz} \cr
H_0 &=& \sum_{\langle ij \rangle \sigma}^{\alpha \beta}
t_{\alpha \beta}^{ij}
 c^{\dagger}_{i \alpha \sigma} c^{~}_{j \beta \sigma} 
 - J_H\sum_i {\bf S}_i.{\mbox {\boldmath $\sigma$}}_i 
+ J\sum_{\langle ij \rangle} {\bf S}_i.{\bf S}_j \cr
&&
~ - \lambda \sum_i {\bf Q}_i.{\mbox {\boldmath $\tau$}}_i
+ {K \over 2} \sum_i {\bf Q}_i^2  
\end{eqnarray}
\noindent
$H_0$ involves two  $e_g$ levels per Mn, with nearest neighbour hopping
$t^{ij}_{\alpha \beta}$ \cite{hop-matr}, and Hunds coupling $J_H$ between the
$e_g$ electrons and the  $t_{2g}$ derived core spins. We assume a 2D square
lattice. The $e_g$ electrons have a coupling $\lambda$ to Jahn-Teller phonons,
while the core spins have an antiferromagnetic (AF) superexchange  $J$  between
them. The stiffness of the phonon modes is $K$, $\mu$ is the chemical potential,
and the magnetic field, ${\bf h} = {\hat z} h$, couples to the core spins.
The `phonons' are
treated in the adiabatic limit and provide a \textit{static}
distortion background for the electrons. The core spins
are also treated 
as classical \cite{class-ref}. At half
doping, $n =0.5$, the clean model has a variety of phases of which we are
interested in the CE-CO-I phase and its response to a magnetic field. Cation
disorder leads to `site disorder' due to the potential generated by the randomly
located rare earth and alkaline earth ions, and `bond disorder' in  $t_{ij}$ and
$J$. We retain only the local random potential since it couples to the electron
density and has a direct impact on charge order. It was recently
shown\cite{sanjeev_kampf} that while off-diagonal disorder is crucial for
obtaining the spin glass phases at half doping, it is diagonal disorder that
plays a dominant role in weakening and eventual disruption of
the CO state. We model the potential $\epsilon_i$ seen at the Mn site as binary 
distribution $P(\epsilon_i) = {1\over 2}(\delta(\epsilon_i - \Delta) +
\delta(\epsilon_i + \Delta))$ with variance  $\Delta^2$. For $H_0$ we use $J_H/t
\rightarrow \infty$, and set 
$\vert {\bf S}_i \vert =1$. $t=1$ is set as the
reference energy scale and we also set $K=1$. We  restrict ourselves to 
$J/t = 0.12$. The  chemical potential $\mu$ is adjusted to keep the electron 
density at $n=1/2$ which is also $x= 1-n =1/2$. 

We use the travelling cluster (TCA) based Monte Carlo method\cite{tca} to solve
the coupled electron-spin-phonon system above.
Classical Monte Carlo techniques have been
extensively used in literature \cite{hyst-mc} to study first order
transitions and the associated hysteresis.
We will explore the dependence of the
melting process on bandwidth $(\lambda/t)$, disorder $(\Delta/t)$, and
temperature $T/t$. 
The  experimental variation of bandwidth (via $r_A$)
implies simultaneous change in $\lambda/t$, $J/t$ and $\Delta/t$, 
while we have varied $\lambda/t$, $J/t$ and $\Delta/t$ independently.
Our primary variation is in $\lambda/t$ and, crudely, we 
compare this to experimental BW variation. We have commented on
this in more detail elsewhere \cite{clean-long}.

 All our results are for cooling
at $h=0$ followed by field sweep. The `upper' and `lower' switching fields,
$h_c^{\pm}$, in the hysteretic response refer to the resistive transitions.

\begin{figure}
\vspace{.2cm}
\centerline{
\includegraphics[width=8.2cm,height=7.6cm,clip=true]{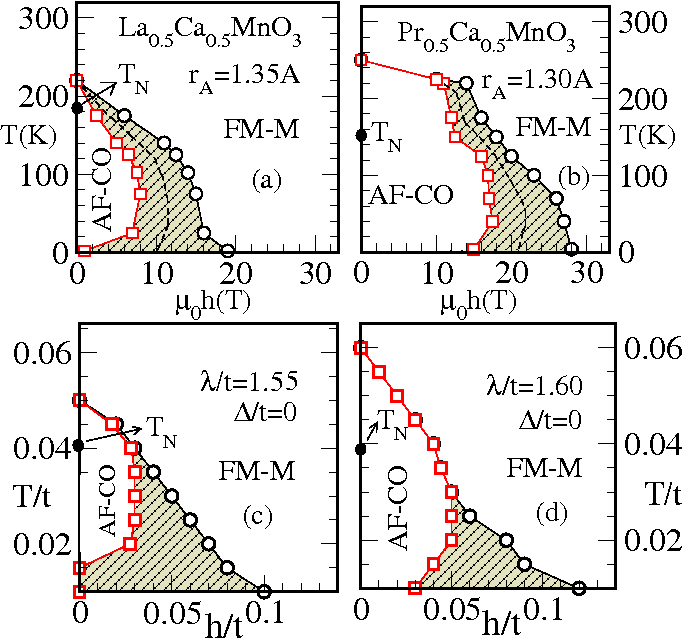}
}
\vspace{.2cm}
\caption{
Colour online:
$h-T$ phase diagrams. (a)-(b) Experimental phase diagrams 
in (a) moderate BW
La$_{0.5}$Ca$_{0.5}$MnO$_3$ and (b) narrow band 
Pr$_{0.5}$Ca$_{0.5}$MnO$_3$.
Both materials have low cation disorder, $\sigma_A \sim 10^{-3}A^2$. 
(c)-(d) Our
results for two values of $\lambda/t$ and  disorder 
$\Delta/t=0$. The parameters
are chosen to mimic  the experimental regimes above 
(see text) in terms of
hysteretic response. Both FM-M and AF-CO-I states 
are locally stable in the
shaded region.}
\end{figure}

The field induced melting of CO  has been studied within mean field 
theory (MFT)
earlier. For a model involving double exchange and Coulomb effects 
Mishra {\it
et al.} \cite{fmlt-th1} showed that suitable parameter choice  could
lead to a low $h_c
\sim 6$T. Fratini {\it et al.} \cite{fmlt-th2}, considered a
more elaborate model and worked out typical $h-T$ phase diagrams 
within MFT, and
Cepas {\it et al.} \cite{cepas} used a variational approach 
at $T=0$ to
estimate~$h_c$. These studies, unfortunately, do not shed 
light on the spatial
character of melting or the impact of disorder. 

Let us first clarify the 
field melting
and hysteresis in manganites with weak cation 
disorder. $\sigma_A$ is  small,
$\sim 10^{-3}A^2$, in the lanthanide (Ln) family Ln$_{0.5}$Ca$_{0.5}$MnO$_3$  
\cite{respaud2000}
so as reference we look at La$_{0.5}$Ca$_{0.5}$MnO$_3$ 
(LCMO), with a relatively
large BW, and Pr$_{0.5}$Ca$_{0.5}$MnO$_3$ (PCMO), 
with a smaller BW.
Fig.1.(a)-(b) reproduces the $h-T$ phase diagrams for 
these compounds from \cite{respaud2000}. At low
temperature  $h_c^+$ is $\sim 20$T for LCMO  and $\sim 30$T 
for PCMO. The field
$h_c^-$, at which CO is {\it recovered} on field 
reduction, is, however, $< 1$T
in LCMO, but $\sim 15$T in PCMO. $h_c^- \sim 0$ at 
low temperature in LCMO indicates
that the FM-M is metastable {\it even at} $h=0$, while 
at smaller BW (PCMO) the
FM-M is no longer metastable at~$h=0$. The $T_{CO}$ 
at $h=0$ for these two
systems are $\sim 220$K and $250$K respectively. 

Panels (c)-(d) show $h-T$ phase
diagrams computed by us at two couplings. For the lower 
value, $\lambda/t=1.55$,
 $T_{CO}/t \sim 0.05$, while for $\lambda/t=1.60$, 
$T_{CO}/t \sim 0.06$.
Using a crude factor of  3/2 to convert from 2D to 
3D and setting $t \sim
0.25$eV, these would correspond to  about 210K and 
250K. More significantly, 
$\lambda/t =1.55$, which is closer to the (FM-M CO-I) phase boundary,
shows $h_c^- =0$, akin
to LCMO, while at $\lambda/t =1.6$, $h_c^-$ is finite and
the FM-M state is no
longer metastable at $h=0$. Although our $h_c^+$  
scales, $\sim 0.1t$, are
larger than in experiments \cite{hc-large}, the typical 
value ${1 \over 2}(h_c^+
+ h_c^-)$ varies roughly similarly in the experiments 
and theory, changing from
10T to 20T between LCMO and PCMO and $\sim 0.05t$ and 
$0.08t$ in Fig.1.(c)-(d).
It is possible that including electron-electron
interaction may reduce the melting fields somewhat \cite{fmlt-th2}
as mean field studies suggest.
The $h-T$ phase diagram is derived from the evolution 
of the  ${\bf q}= \{0,0\}$
feature in the magnetic structure factor $S({\bf q })$ 
and the `volume fraction'
$V_{co}$ \cite{vol-frac}  of the CO phase. 
\begin{figure}
\centerline{
\includegraphics[width=8.0cm,height=7.8cm,clip=true]{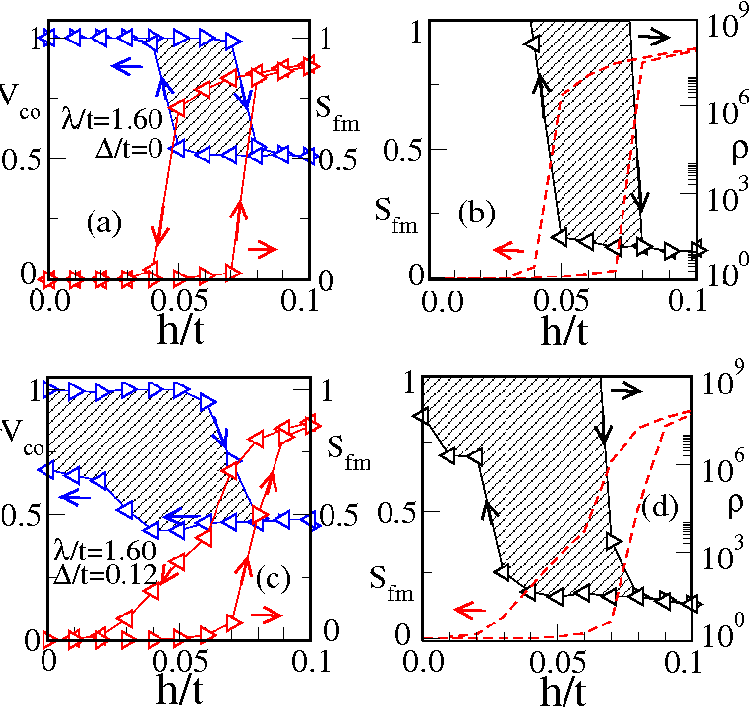}
}
\vspace{.2cm}
\caption{
Colour online: Response to field cycling at
$\lambda/t=1.6$ and $T/t=0.02$.
(a)-(b) Clean system  $\Delta/t=0$: (a) field
response of the CO volume fraction
(see text) and the FM structure factor $S_{fm} $,
(b) the resistivity $\rho(h)$.
(c)-(d) Corresponding results in the disordered case
$\Delta/t=0.12$. Shaded
region is the hysteretic window. Contrast the abrupt
switching in the clean limit with the broad crossover in
the disordered case.
Note, both the CO volume fraction and the FM magnetic
structure factors are
normalized to unity.}
\end{figure}
\begin{figure}[t]
\centerline{
\includegraphics[width=8.2cm,height=8.0cm,clip=true]{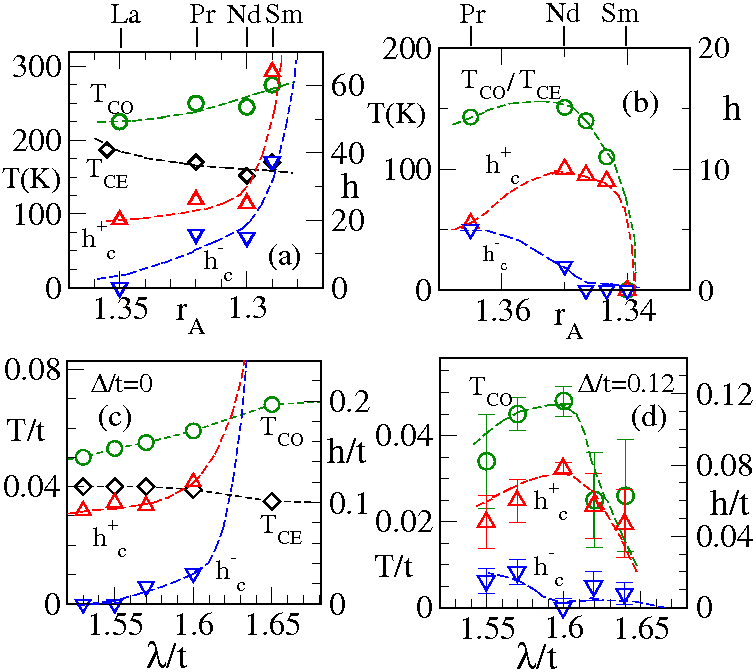}
}
\vspace{.2cm}
\caption{
Colour online: $T_{CO}$, and $h_c^{\pm}$ at low $T$ - comparing 
experiment and
theory. (a)-(b) Data from experiments: (a) the 
Ln$_{0.5}$Ca$_{0.5}$MnO$_3$
family, with typical $\sigma_A \sim 10^{-3}A^2 $, ~(b) the
Ln$_{0.5}$Sr$_{0.5}$MnO$_3$ family with typical $\sigma_A 
\sim 10^{-2}A^2$.
Notice the rapid increase in $h_c^{\pm}$ with decreasing 
$r_A$  at $r_A \sim
1.29~A$ in (a), and the {\it collapse} of $h_c^{\pm}$ 
with decreasing $r_A$
at $r_A \sim 1.34$ in (b). (c)-(d) Our results on the 
$\lambda/t$ dependence of
$T_{CO}$, $T_{CE}$ and $h_c^{\pm}$. (c) Clean limit 
$\Delta/t=0$,  (d)
disordered systems, $\Delta/t=0.12 $. The lines are a 
guide to the eye. }
\end{figure}
\begin{figure}[t]
\vspace{.2cm}
\centerline{
\includegraphics[width=8.0cm,height=5.6cm,clip=true]{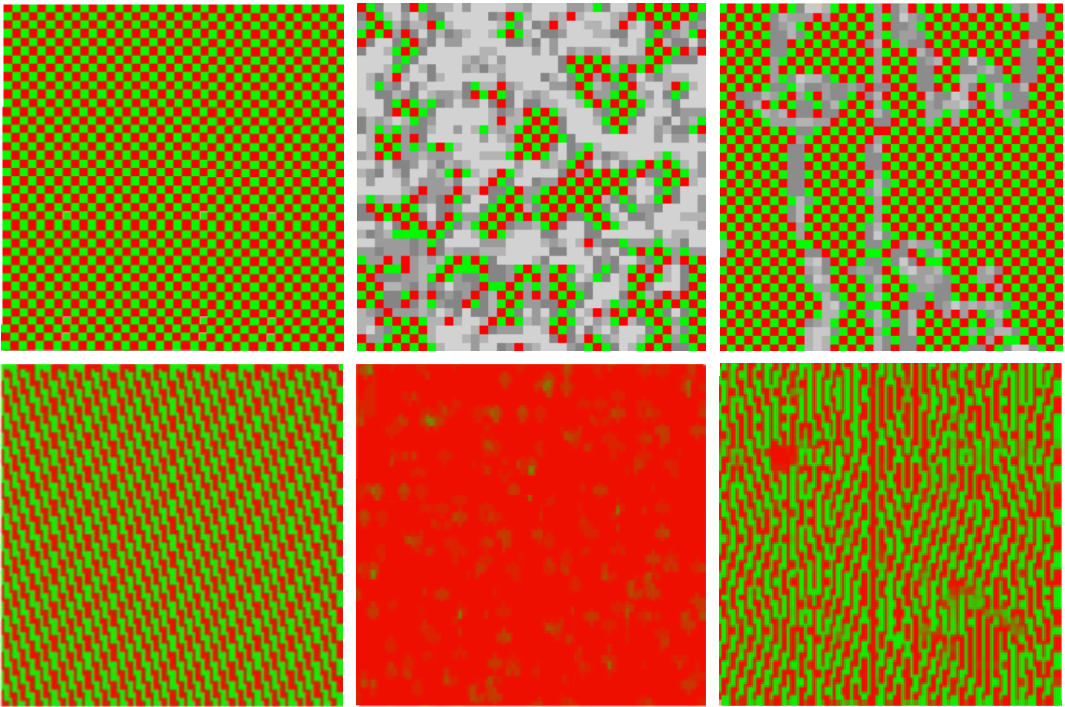}
}
\vspace{.4cm}
\caption{
Colour online: Spatial snapshots of local charge order (top row) 
and nearest neighbour magnetic correlation $ {\bf S}_i.{\bf S}_j$ 
(bottom row)
on field cycling, $\lambda/t=1.55$, $\Delta/t=0$, 
$T/t=0.02$. $h=0$ (left)
initial state,  $h > h_c^+$ (center), and $h=0$ (right) after 
the field sweep. 
The lattice size is $40 \times 40$.  In the bottom panel the 
red bonds are FM
the green bonds are AF. In the top panel grey regions 
imply metallic regions and
the red and green regions are the $0.5-\delta$ and 
$0.5+\delta$ onsite charge
densities.}
\vspace{.2cm}
\end{figure}

Fig.2 shows 
these features at 
$\lambda/t=1.6$ and $T/t=0.02$. 
Fig.2.(a) shows the `switching'  of $V_{co}$ and  
$S_{fm} = S(0,0)$ on field
cycling, for a clean system. The complementary switching 
in FM and CO volume
fraction is sharp and simultaneous, along with the
transition in resistivity
$\rho(h)$ shown in Fig.2.(b). Notice the {\it residual $V_{co}$ } 
in the field
melted metallic state in Fig.2.(a). For a weakly 
disordered CO state, at
$\Delta/t=0.12$, Fig.2.(c) shows that the CO melts on 
increasing $h$, but fails
to recover on downward sweep. The magnetism in Fig.2.(c) 
and $\rho(h)$ in
Fig.2.(d) still show hysteresis  but the  transitions 
are quite broad. This
suggests that the field induced `melting' in the disordered 
case {\it occurs gradually}, 
unlike the  abrupt switching in the clean case. 

Using indicators as
shown in Fig.1 and Fig.2 we made a comprehensive comparison 
between experiment
and theory for the dependence of transition temperatures 
and low $T$ melting
fields on BW and disorder. Fig.3.(a) is  data from the 
`low disorder'
Ln$_{0.5}$Ca$_{0.5}$MnO$_3$ \cite{respaud2000} family, 
and Fig.3.(b), is for the
moderate disorder Ln$_{0.5}$Sr$_{0.5}$MnO$_3$ 
\cite{tokura96} family. We compare
these with our results on the $\lambda/t$ dependence 
of $T_{CO}$ and $h_c^{\pm}$
in Fig.3.(c) and Fig.3.(d), with $\Delta/t=0$ and 
$\Delta/t=0.12$ respectively.

({\bf A})~In the clean limit, Fig.3.(a), $T_{CO}$ 
increases with decreasing BW,
$T_{CE}$ decreases, and the fields $h_c^{\pm}$ 
{\it increase rapidly} beyond 
Sm$_{0.5}$Ca$_{0.5}$MnO$_3$ (SCMO). A comparison of 
Fig.3.(a) with Fig.3.(c)
shows that our results capture all the experimental 
trends in the clean limit.
The increase in $T_{CO}$ is due to the decrease 
in kinetic energy with
decreasing $r_A$ (increasing $\lambda/t$ in our 
case), while $T_{CE}$ is
suppressed because CE magnetic order is also kinetic energy 
driven and gives way
to a `G type' $\{ \pi, \pi\}$ phase at large 
$\lambda/t$. The increasing
separation of $T_{CO}$ and $T_{CE}$ implies that the 
charge order becomes 
progressively independent of CE magnetic order with 
decreasing BW, and so {\it
less responsive to an applied field.} This is precisely 
what one observes for
$h_c^{\pm}$  in Fig.3.(a) \& (c), where below a 
critical BW the CO no longer
melts! In fact for $\lambda/t > 1.60$ we observe 
that the high field state
is FM-CO and not FM-M. The magnetic phase change no longer 
drives a resistive
transition. 

({\bf B})~For the disordered family, 
Fig.3.(b), $T_{CO}$  and
$h_c^{\pm}$ {\it decrease rapidly} as we head towards 
small BW
Sm$_{0.5}$Sr$_{0.5}$MnO$_3$ (SSMO). Our results in Fig.3.(d) 
probe BW variation
at disorder $\Delta/t =0.12$, and capture the correct 
dependence. The fall in
$T_{CO}$ at large $\lambda/t$ is due to the competition 
between the `stiffness'
$(K_{eff}(\lambda)$, say) of the CO state, which weakens 
at large $\lambda/t$
\cite{dis-long}, and the pinning effect of the random 
potential $\epsilon_i$.
This is similar to the random field Ising model 
\cite{rfim-ref} where the
exchange $(J_{eff})$ serves the role of our 
`stiffness' and the random field
$(h_i)$ is akin to $\epsilon_i$. As is well known, 
long range order is lost
below a threshold for $J_{eff}/\sqrt{ \langle h_i^2 \rangle}$, 
equivalent to our
$K_{eff}/\Delta$. The weakening and ultimate
destruction of the zero field CO state with 
decreasing $r_A$ (or $t/\lambda$)
automatically suppresses the melting fields 
$h_c^{\pm}$. The non monotonic
character of $T_{CO}$ and $h_c$, with a 
reduction also at the large BW end, is
due to the presence of the competing FM-M phase~at~$\lambda/t 
< 1.5$.
\begin{figure}
\vspace{.2cm}
\centerline{
\includegraphics[width=6.5cm,height=5.5cm,clip=true]{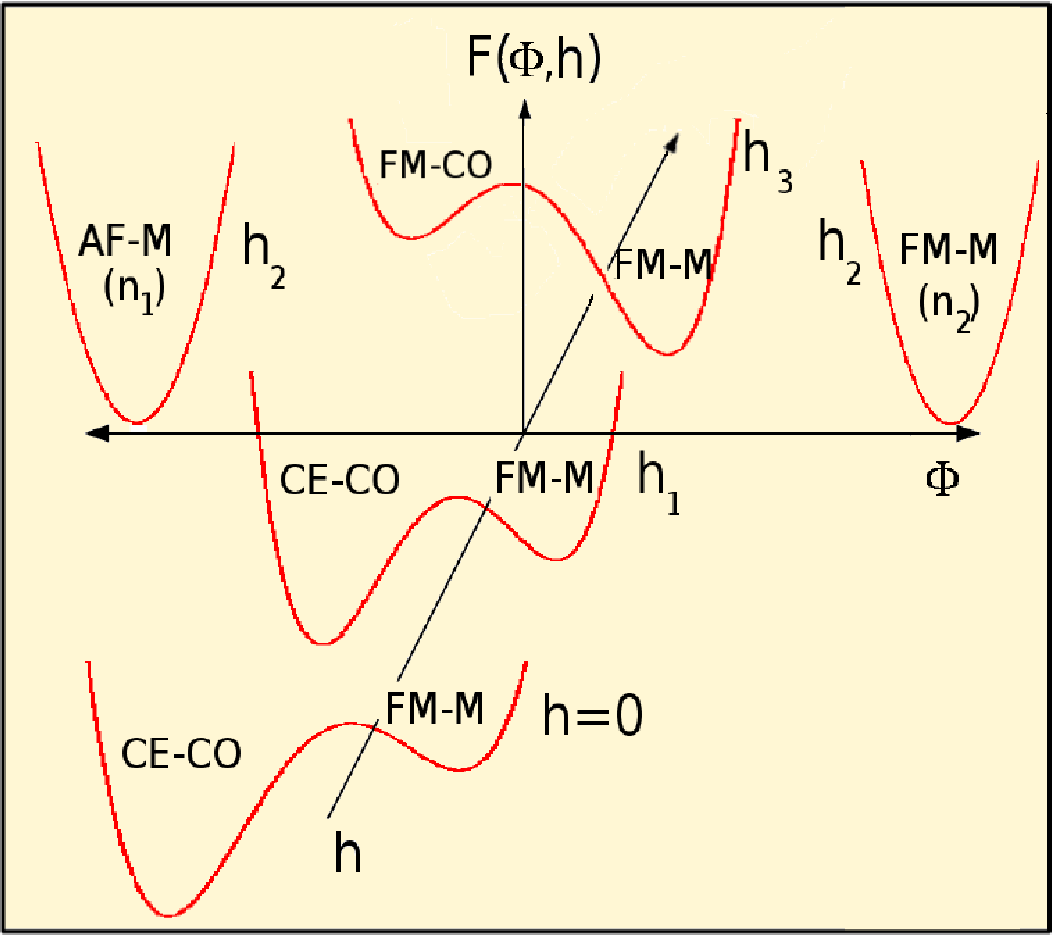}
}
\vspace{.4cm}
\caption{
Colour online: Landau landscape. We plot the free energy $F(\phi, h)$, where
$\phi$ is the `order parameter' (on the $x$ axis) and can have FM-M, CE-CO-I,
AF-M, or AF-CO character. $h$ increases from $h_1 $ to $h_3$(on the $y$ axis).
We have not shown
the metastable FM-CO state, for $h \sim h_2$, to avoid cluttering.}
\vspace{.2cm}
\end{figure}

While the trends are well captured, let us compare the experimental $T_{CO}$
scales with theory. 
At the `large' BW end, $r_A \sim 1.36A$, `clean' LCMO has $T_{CO} \sim 220$K,
while `disordered' Pr$_{0.5}$Sr$_{0.5}$MnO$_3$ (PSMO) has a suppressed $T_{CO}
\sim 150$K. If we identify $\lambda/t \sim 1.55$ crudely as the equivalent, the
clean $T_{CO}$ is $\sim 0.055t \sim 225$K, and the `disordered' $T_{CO}$ is
$\sim 0.035t \sim 145$K. The peak $T_{CO}$ is $ \sim 160$K in Fig.3.(b) and it
is $\sim 0.05t \sim 200$K in theory. Given that we are solving a 2D model and
have not fine tuned parameters the correspondence is reasonable. 

The
thermodynamic indicators, however, do not reveal the spatial character of the
`melted' state, while Fig.2.(a) \& (c)  suggest something unusual: there is
finite CO phase fraction in the `FM-M'! Our map of the density field in Fig.4
reveals how the system is spatially organised. The lower panels show that the CE
phase transforms to a FM at large field (and only partially recovers CE  order
on  reverse sweep) but the large $h$ density field is {\it strongly
inhomogeneous}. The middle panel, top row, shows charge modulated `minority'
regions in a homogeneous background, surviving  way past the CE $\rightarrow$ FM
transition. For the parameter values in Fig.4, and overall for $\lambda/t
< 1.6$, the `high field' state is a {\it percolative} metal. Beyond
$\lambda/t = 1.6$ it is an insulating FM-CO. 

The  peculiar high field state
requires us to examine the various phases that `compete' in the manganites
around $x \sim 0.5$ at finite $h$. At $h=0$ the $x=0.5$ CE-CO-I is separated
from a FM-M phase at lower $x$, and an AF-M phase at larger $x$, by windows of
PS \cite{kp-am-pm1}. However, beyond a small threshold field a homogeneous 
$x=0.5$ state is no longer possible \cite{clean-long} and the $x=0.5$ system 
breaks up into either (a)~$x <  0.5$ FM-M and  $x >  0.5$ AF-M patches, for
$\lambda/t <  1.6$, or (b)~$x <  0.5$ FM-CO and  $x >  0.5$ AF-M patches for
$\lambda/t >  1.6$. These intermediate field states evolve, respectively, into a
homogeneous FM-M or FM-CO state at large $h$. While the finite $h$ {\it
equilibrium state} is complicated by PS, field sweep experiments (and our
simulation) also probe the existence of metastable states in which the system
can get trapped. Fig.5 summarises the situation at $\lambda/t \sim  1.55$ in
terms of a Landau free energy schematic. At $h=0$ the CE-CO-I is the absolute
minimum while the FM-M is metastable (ref Fig.1.(c)). This continues to finite
$h$, we show a typical field $h=h_1$. At $h=h_2(\lambda)$ the $x=0.5$ state 
phase separates  
into AF-M and FM-M, and there is also a metastable FM-CO. Finally, at some $h
= h_3(\lambda)$ the equilibrium state is a homogeneous FM-M but 
the continuing presence of the low energy
metastable FM-CO state 
would affect all low $T$ field sweep experiments. 

To conclude, we have clarified how non-equilibrium effects and
disorder control the field driven conductance switching and insulator-metal
transition in the half doped manganites. We discover that the field induced
conducting state is spatially inhomogeneous, even in the absence of quenched
disorder. Our results on the $h-T$ phase diagram and the spatial
character of the melted state address existing results and growing body of
spatially resolved data.

We acknowledge use of the Beowulf cluster at HRI.

{}

\begin{thebibliography}{99}
\bibitem{hop-matr} 
The detailed orbital and $ij$ dependence is given in \cite{kp-am-pm1}.
\bibitem{vol-frac}
Our CO volume fraction is a measure of the local CO
correlation, obtained by essentially counting the number of sites
on the lattice which are correctly coordinated to form a
local $(\pi,\pi)$ charge order.
We prefer this `short range' measure since it is useful in
both 2D and 3D, and correlates with the transport.
Features depending on long range order will not survive
the effect of disorder in 2D.
\end{thebibliography}
\end{document}